\documentclass[10pt,twocolumn,letterpaper]{article}

\usepackage{cvpr}              
\usepackage{multirow}
\usepackage{multicol}
\usepackage{setspace}

\usepackage[dvipsnames]{xcolor}

\usepackage{ifpdf}
\ifpdf
\else
\fi

\usepackage{color}
\definecolor{cyan}{cmyk}{1,0,0,0}
\definecolor{darkgreen}{rgb}{0,0.5,0}
\definecolor{orange}{rgb}{1,0.5,0}
\definecolor{magenta}{cmyk}{0,1,0,0}
\definecolor{darkyellow}{cmyk}{0,0,0.75,0}
\definecolor{gray}{rgb}{0.8,0.8,0.8}
\usepackage{multirow}
\usepackage{booktabs}
\usepackage{array}
\usepackage{setspace}
\usepackage{makecell}
\usepackage{placeins}
\definecolor{cvprblue}{rgb}{0.21,0.49,0.74}
\usepackage[pagebackref,breaklinks,colorlinks,citecolor=cvprblue]{hyperref}
\usepackage{colortbl,xcolor,array}

\def\confName{CVPR}
\def\confYear{2024}
\def\authorBlock{
	Jie Lian$^{1} $ \qquad Lizhi Wang$^{1}$ \qquad Lin Zhu$^{1}$ \qquad Renwei Dian$^{2}$ \qquad Zhiwei Xiong$^{3}$ \qquad Hua Huang$^{4}$ \\
	$^{1}$Beijing Institute of Technology \qquad $^{2}$Hunan University \\
	$^{3}$University of Science and Technology of China  \qquad $^{4}$Beijing Normal University \\}

\newif\ifreview 
\newif\ifarxiv 
\newif\ifcamera 
\newif\ifrebuttal

\title{Physics-Inspired Degradation Models for Hyperspectral Image Fusion}

\begin{document}
\author{\authorBlock}
\maketitle
\begin{abstract}
The fusion of a low-spatial-resolution hyperspectral image (LR-HSI) with a high-spatial-resolution multispectral image (HR-MSI) has garnered increasing research interest. However, most fusion methods solely focus on the fusion algorithm itself and overlook the degradation models, which results in unsatisfactory performance in practical scenarios. To fill this gap, we propose physics-inspired degradation models (PIDM) to model the degradation of LR-HSI and HR-MSI, which comprises a spatial degradation network (SpaDN) and a spectral degradation network (SpeDN). SpaDN and SpeDN are designed based on two insights. First, we employ spatial warping and spectral modulation operations to simulate lens aberrations, thereby introducing non-uniformity into the spatial and spectral degradation processes. Second, we utilize asymmetric downsampling and parallel downsampling operations to separately reduce the spatial and spectral resolutions of the images, thus ensuring the matching of spatial and spectral degradation processes with specific physical characteristics. Once SpaDN and SpeDN are established, we adopt a self-supervised training strategy to optimize the network parameters and provide a plug-and-play solution for fusion methods. Comprehensive experiments demonstrate that our proposed PIDM can boost the fusion performance of existing fusion methods in practical scenarios. 
\end{abstract}

\section{Introduction}
\label{sec:intro}
\hspace{1em} The hyperspectral image (HSI) record the power distribution of a scene in three-dimensional data, which delineates the spectral intensity across varied wavelengths for each pixel location \cite{hsi, hsi2, hsi3, hsi4, hsi5, hsi6, hsi7}. However, hyperspectral imaging system frequently confronts a trade-off between spatial and spectral resolutions due to hardware restrictions \cite{trade-off1, trade-off2}. Consequently, HSI commonly exhibits low spatial resolution, which limits its range of applications \cite{hsi-app1, hsi-app2}. Fortunately, a high-spatial-resolution HSI (HR-HSI) can be obtained by fusing a low-spatial-resolution HSI (LR-HSI) with a high-spatial-resolution MSI (HR-MSI) \cite{fusion-based1, fusion-based2, fusion-based3, fusion-based4, fusion-based5, fusion-based6, fusion-based7, pan2, pan3}, which breaks the hardware restrictions and demonstrates promising application prospects \cite{app-pro1, app-pro2, app-pro3, app-pro4, app-pro5}.

\begin{figure}[!t]
	\setlength{\abovecaptionskip}{-2pt}
	\setlength{\belowcaptionskip}{0pt}
	\begin{center}
		\includegraphics[scale=0.27]{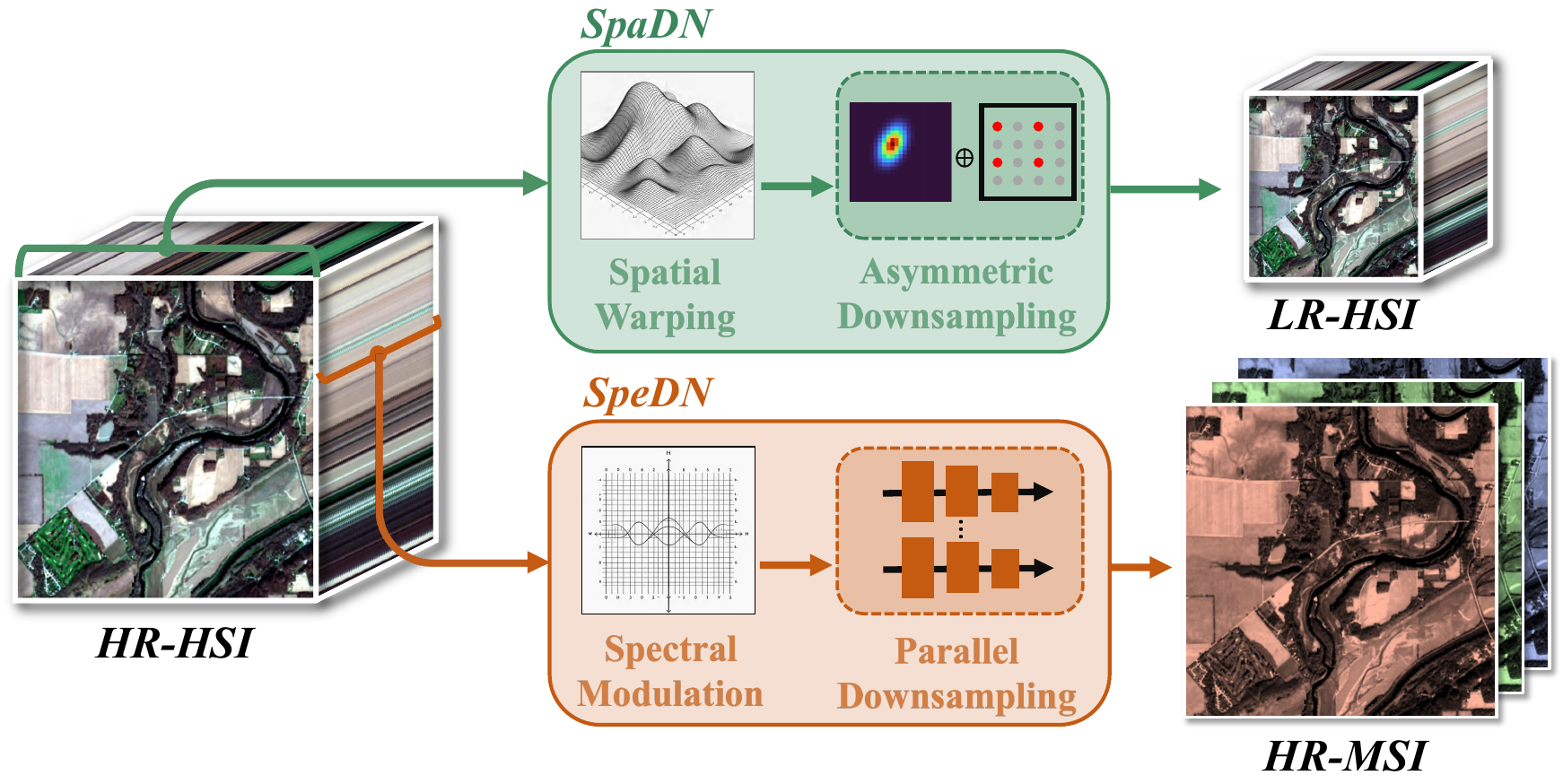}
	\end{center}
	\caption{Illustration of our proposed PIDM. We utilize SpaDN and SpeDN to represent the spatial and spectral degradation models of LR-HSI and HR-MSI, respectively.}
	\label{fig:intro}
\end{figure}

While the concept of fusing LR-HSI with HR-MSI holds great promise, it is imperative to recognize that the success of such fusion techniques is not solely dependent on the fusion algorithm itself. Equally important is the accurate modeling of degradation processes in various imaging systems. The essence of fusion is to inverse the degradation process \cite{inverse}, which makes accurate degradation models crucial for achieving optimal fusion results. Regrettably, most fusion methods predominantly focus on the fusion algorithms and overlook the degradation models. Such oversight may compromise the performance of these fusion methods in practical scenarios, as these methods would not cope with the degradation mechanisms in the hyperspectral imaging process.

In this paper, we propose physics-inspired degradation models (PIDM) for hyperspectral image fusion, as shown in Fig. \ref{fig:intro}. We observe that the lens in the hyperspectral imaging system may manifest aberrations \cite{uniform1, uniform2}, such as distortion and blur, which result in varying degrees of degradation for pixels at different positions within an image. Moreover, hyperspectral image degradation corresponds to a complex physical process that necessitates the consideration of various physical characteristics rather than solely focusing on adjustments to the image dimensions. Hence, we propose PIDM to characterize the degradation processes of LR-HSI and HR-MSI, which comprises a spatial degradation network (SpaDN) and a spectral degradation network (SpeDN). SpaDN and SpeDN are designed based on two insights. First, we employ spatial warping and spectral modulation operations to simulate lens aberrations, thereby introducing non-uniformity into the spatial and spectral degradation processes. Second, we utilize asymmetric downsampling and parallel downsampling operations to separately reduce the spatial and spectral resolutions of the images, thus ensuring the matching of spatial and spectral degradation processes with specific physical characteristics.

Once SpaDN and SpeDN are established, we adopt a self-supervised training strategy to optimize the network parameters. Specifically, we employ SpaDN to spatially degrade the HR-MSI and SpeDN to spectrally degrade the LR-HSI, followed by ensuring the degraded images maintain a consistent appearance through a loss function. Our proposed PIDM can seamlessly integrate into various existing fusion methods and provides a plug-and-play solution, which exhibits a remarkable level of flexibility. We conduct comprehensive experiments on four real data using six fusion methods. Experimental results consistently show that when our PIDM are equipped, these fusion methods can achieve boosted fusion results. 

The main contributions of our work are summarized as follows:

\begin{enumerate}[leftmargin=2.4em]
	\item We precisely model the degradation processes of LR-HSI and HR-MSI inspired by the physical realities and provide a plug-and-play solution for fusion methods.
	\item We introduce non-uniformity, as well as asymmetric and parallel downsampling, into the LR-HSI and HR-MSI degradation models while optimizing the network parameters of the proposed models through a self-supervised strategy. 
	\item We conduct comprehensive experiments on real and synthetic data to demonstrate the validity and authenticity of our proposed physics-inspired degradation models. 
\end{enumerate}

\section{Related Work}
\label{sec:relatedwork}
\hspace{1em} In the HSI fusion field, researchers have conducted substantial work on building fusion algorithms \cite{fusion1, fusion2, fusion3, fusion4, fusion5, fusion6, fusion7, fusion8, fusion9} but relatively little on estimating degradation models. In this section, we provide a concise overview of previous efforts \cite{hypconet, miae, dfmf, zsl, daem, busifusion} in degradation modeling. For related work on the fusion algorithms, please refer to the review literature \cite{review1, review2, review3, review4, review5}.

The hyperspectral coupled network (HyCoNet) \cite{hypconet} adopts two trainable matrices to represent the spatial and spectral degradation models and optimize the parameters of these matrices in an unmixing framework \cite{unmixing1, unmixing2}. Similarly, the model inspired autoencoder (MIAE) \cite{miae} proposes a blind estimation network to estimate degradation models, which employs two trainable matrices to capture the LR-HSI and HR-MSI degradation processes. The deep framework for multi-satellite HSI fusion (DFMF) method \cite{dfmf} presents degradation models for the LR-HSI and HR-MSI under the conditions of different imaging satellites, with two weight tensors proposed to bridge the gap of spectral drifts between the images from different satellites. The zero-shot learning (ZSL) \cite{zsl} method proposes an iterative optimization framework to quantitatively estimate the spatial and spectral responses of imaging sensors, which aims to model degradation progress in an explicit way. DAEM \cite{daem} establishes an explicit estimation method to model the spatial and spectral degradation processes based on Gaussian theory, which focuses on reducing the estimation parameters of the degradation matrices. The blind unsupervised single image fusion (BUSIF) \cite{busifusion} method estimates the degradation models through a neural representation, which adopts multilayer perception networks to generate the degradation matrices. 

The methods above presume that spatial and spectral degradations are uniform across the spatial dimension. However, this assumption does not correspond to the physical reality of hyperspectral imaging systems, which limits their effectiveness in practical scenarios. Unlike these methods, we introduce non-uniformity into degradation progress and use two designed downsampling operations to ensure the matching of degradation progress with specific physical characteristics.

\begin{figure}[!t]
	\begin{center}
		\hspace*{-0.5cm}
		\includegraphics[scale=0.185]{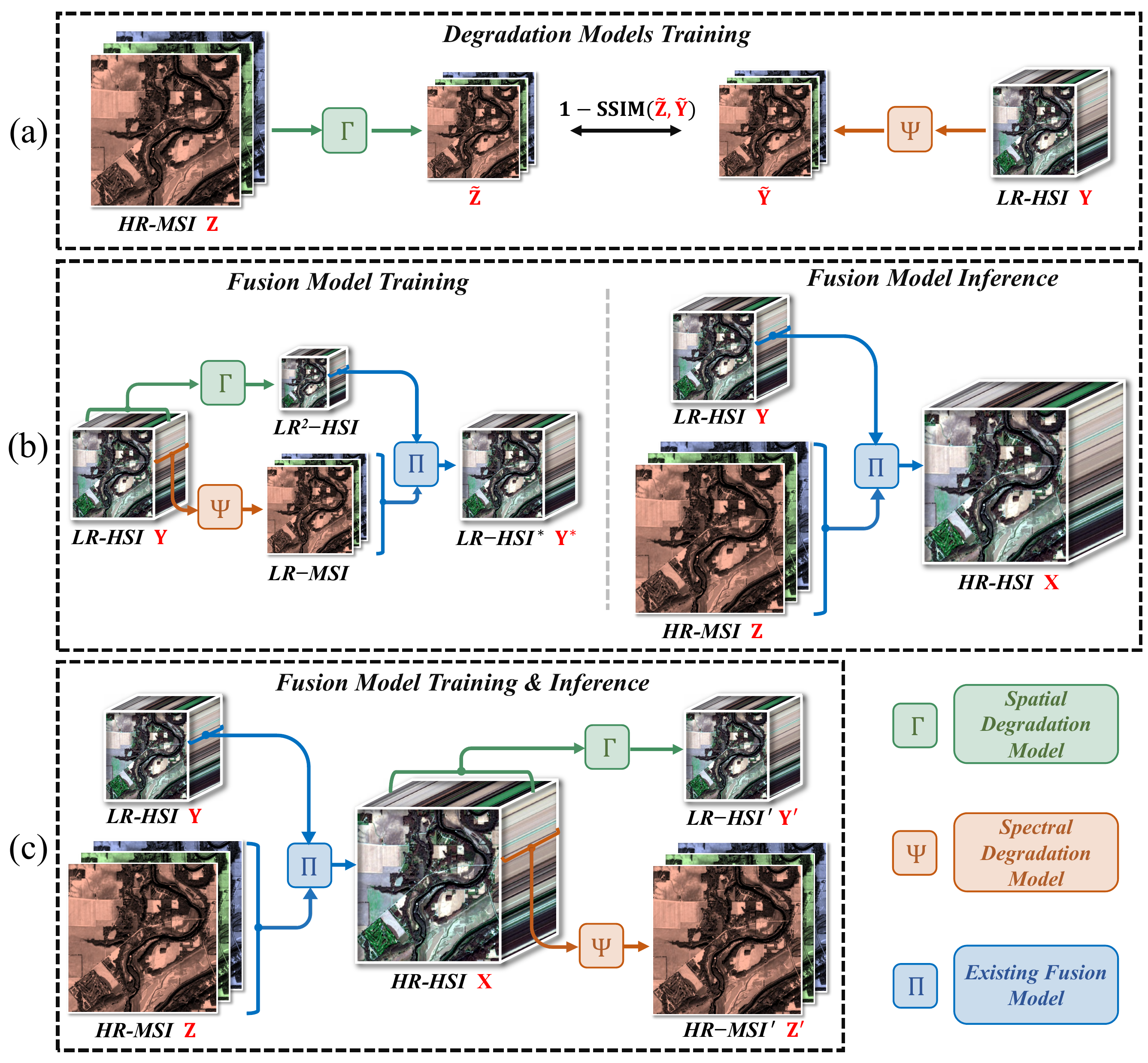}
	\end{center}
	\caption{The training strategy and usage modes of degradation models. (a) Self-supervised training strategy. (b) Training samples generation mode. (c) Objective function construction mode.}
	\label{fig:mode}
\end{figure}

\section{Method}
\label{sec:method}
\hspace{1em} In this paper, we propose two networks, SpaDN and SpeDN, to model the LR-HSI and HR-MSI degradation process. This section initially presents the overall framework of our approach and subsequently provides the implementation details of SpaDN and SpeDN, respectively. 

\subsection{Overall  Framework}
\hspace{1em} The content of this subsection is arranged as follows. We start with a mathematical description of the degradation networks, followed by explanations of the training and application methods.

Given an HR-HSI $\textbf{X} \in \mathbb{R}^{ H \times W \times C}$, where  $H$, $W$, and $C$  represents the height, width, and the number of spectral bands of the image, respectively. The LR-HSI $\textbf{Y} \in \mathbb{R}^{h \times w \times C}$  is produced through the spatial degradation of the HR-HSI $\textbf{X}$:
\begin{equation}
	\textbf{Y} = \Gamma(\textbf{X})
	\label{eq:eq1}
\end{equation}
where $\Gamma$ represents the spatial degradation model. Compared with the HR-HSI $\textbf{X}$, the LR-HSI $\textbf{Y}$ shares equivalent spectral resolution but possesses lower spatial resolution ($H \gg h, W \gg w$). Moreover, the HR-MSI $\textbf{Z} \in \mathbb{R}^{H \times W \times c}$ is produced through the spectral degradation of the HR-HSI $\textbf{X}$:
\begin{equation}
	\textbf{Z} = \Psi(\textbf{X})
	\label{eq:eq2}
\end{equation}
where $\Psi$ represents the spectral degradation model. Compared with the HR-HSI $\textbf{X}$, the HR-MSI $\textbf{Z}$ maintains the same spatial resolution but exhibits reduced spectral resolution ($C \gg c$). 

In this paper, we utilize SpaDN and SpeDN to represent the spatial degradation model $\Gamma$ and the spectral degradation model $\Psi$, respectively. As illustrated in Fig. \ref{fig:mode} (a), we employ a self-supervised strategy \cite{self} to optimize the network parameters of SpaDN and SpeDN as:
\begin{equation}
	\widetilde{\textbf{Z}}  = \text{SpaDN}(\textbf{Z}) \leftrightarrow \text{SpeDN}(\textbf{Y}) = \widetilde{\textbf{Y}}
	\label{eq:eq3}
\end{equation}
where $\widetilde{\textbf{Z}} \in \mathbb{R}^{h \times w \times c}$ and $\widetilde{\textbf{Y}} \in \mathbb{R}^{h \times w \times c}$ represents the spatially degraded HR-MSI and the spectrally degraded LR-HSI, respectively. As the training progresses, the spatially degraded HR-MSI $\widetilde{\textbf{Z}}$ should grow closer the spectrally degraded LR-HSI $\widetilde{\textbf{Y}}$. We adopt the structural similarity index (SSIM) \cite{ssim} loss to capture the perceptual differences between the spatially degraded HR-MSI $\widetilde{\textbf{Z}}$ and the spectrally degraded LR-HSI $\widetilde{\textbf{Y}}$:
\begin{equation}
	\mathcal L(\theta ) = 1 - \text{SSIM}(\widetilde{\textbf{Z}}, \widetilde{\textbf{Y}})
	\label{eq:eq4}
\end{equation}
where $\theta$ represents the network parameters of SpaDN and SpeDN. Once training is completed, SpaDN and SpeDN are integrated into fusion methods for practical utilization.

As shown in Fig. \ref{fig:mode} (b-c), the utilization of degradation models can be roughly categorized into two modes, namely, training samples generation mode (TSG-mode) and objective function construction mode (OFC-mode). In TSG-mode, degradation models are used to generate training samples for training the fusion model $\Pi$ during the training phase. The training loss function is set to $\mathcal{L}_{\text{TSG-mode}} =  \| \textbf{Y} - \textbf{Y}^{*}\|_1$, where $\textbf{Y}^{*}$ represents the fused LR-HSI obtained by the fusion model $\Pi$. In the inference phase, the trained fusion model $\Pi$ is directly employed to fuse the LR-HSI $\textbf{Y}$ and the HR-MSI $\textbf{Z}$ to obtain the target HR-HSI $\textbf{X}$. In OFC-mode, degradation models are used for the design of objective function:
\begin{equation}
	\begin{aligned}
		\mathcal{L}_{\text{OFC-mode}} &=  \| \textbf{Y} - \textbf{Y}'\|_1 + \| \textbf{Z} - \textbf{Z}'\|_1 \\
		&= \| \textbf{Y} - \Gamma(\textbf{X})\|_1 +  \| \textbf{Z} - \Psi(\textbf{X})\|_1.
	\end{aligned} 
\end{equation}
where $\textbf{X}$ represents the target HR-HSI, $\textbf{Y} '$ and $\textbf{Z}'$ represent the  degenerated images of the target HR-HSI $\textbf{X}$, respectively. In OFC-mode, once the training is completed, the final fusion result is determined. From the above analysis, it can be observed that the establishment and training of the fusion model rely heavily on the degradation models.

\begin{figure*}[!t]
	\setlength{\abovecaptionskip}{3pt}
	\setlength{\belowcaptionskip}{-0pt}
	\begin{center}
		\hspace*{-0.3cm}
		\includegraphics[scale=0.4]{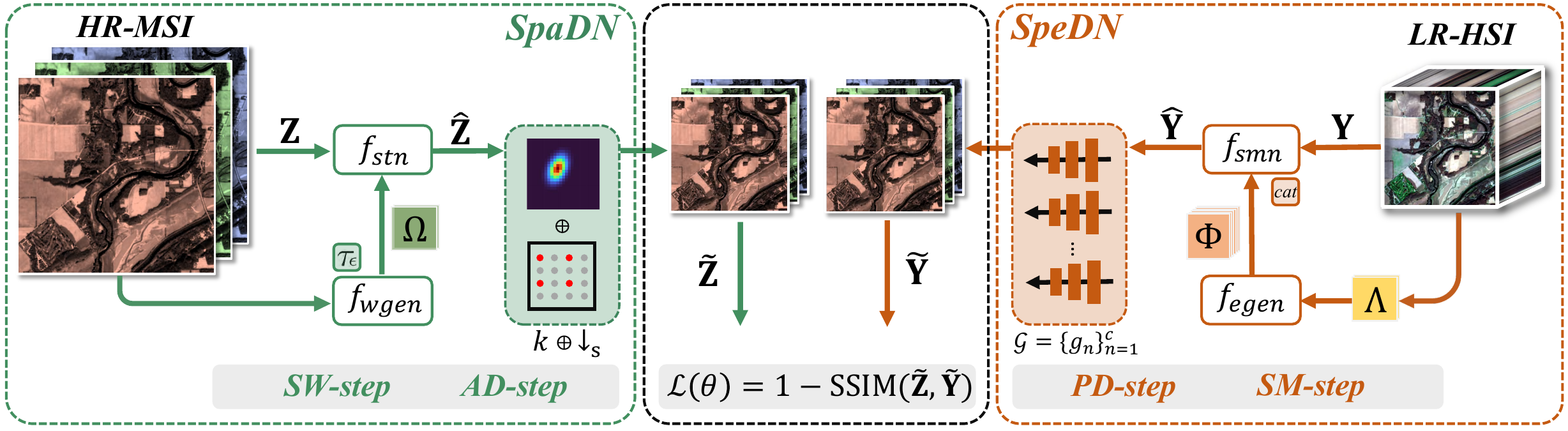}
	\end{center}
	\caption{Framework of proposed PIDM. The structures of our designed SpaDN and SpeDN are mirror-symmetric. The implementation of SpaDN is divided into two SW-step and AD-step, and the implementation of SpeDN is divided into two SM-step and PD-step. The network parameters are optimized using a self-supervised training strategy.}
	\label{fig:pipeline}
\end{figure*}

\subsection{Spatial Degradation Network}
\hspace{1em} SpaDN performs spatial resolution reduction on the image, which represents a physical process of transforming the image from high to low spatial resolution, with the overall process exhibiting non-uniformity. 
As illustrated in Fig. \ref{fig:pipeline}. the SpaDN divides this process into two steps for implementation, namely, the spatial warping step (SW-step) and the asymmetric downsampling step (AD-step). Next, we introduce the details of SW-step and AD-step, respectively.

\textbf{\textit{SW-step:}} The chief goal of the SW-step is to simulate lens aberrations, thereby introducing non-uniformity into the spatial degradation processes. To achieve this goal, we employ a spatial warping operation as:
\begin{equation}
	\widehat{\textbf{Z}} =f_{stn}(\mathcal{\textbf{Z}}, \Omega)
	\label{eq:eq5}
\end{equation}
where $\Omega \in \mathbb{R}^{ H \times W \times 2}$ represents the warping field, which stores the degree and directions of pixel deformations across various locations, with the directions defined along the vertical (y-axis) and horizontal (x-axis) planes\footnote{In the coordinate system of this paper, the y-axis represents image height, and the x-axis represents image width, with positive y pointing downward and positive x pointing rightward. We use the coordinate format of $(y, x)$ to better align with the (height, width) notation in the data.}. $f_{stn}$ represents the spatial transformer network \cite{stn} employed to warp the HR-MSI $\textbf{Z}$, and $\widehat{\textbf{Z}} \in \mathbb{R}^{H \times W \times c}$ is the warped HR-MSI.

We encode position information into the warping field $\Omega$, which then guides the spatial transformer networks $f_{stn}$ to perform diverse spatial warping on the pixels of the HR-MSI $\textbf{Z}$ at various locations. Next, we provide the specific formulation for generating the warping field $\Omega:$

\begin{equation}
	\begin{aligned}
		\Omega = \tau_\epsilon(f_{wgen}(\textbf{Z}))
	\end{aligned}
	\label{eq:eq6}
\end{equation}
where $f_{wgen}$ represents the warping field generator. This generator is built upon a convolutional neural network architecture similar to UNet \cite{unet}, which consists of encoder and decoder sections with skip connections. Both the encoder and decoder are composed of stacked convolutional layers and ReLU activation layers. 

Furthermore, we introduce a truncation function $\tau_\epsilon$ to limit the values within the warping field $\Omega$. The function is defined as:
\begin{equation}
	\tau_\epsilon(t) = 
	\begin{cases} 
		t, & \text{if } |t| \leq \epsilon \\
		0, & \text{if } |t| > \epsilon 
	\end{cases}
	\label{eq:eq7}
\end{equation}
where $\epsilon$ represents the threshold of the truncation function $\tau_\epsilon$. This procedure becomes crucial when outliers exist in the warping field $\Omega$. The outliers might emerge due to errors, noise, or other unexpected factors, which leads to illogical deformations in the image. Setting these outliers to zero through the truncation function $\tau_\epsilon$, thereby preventing these outliers from adversely affecting the warping results. 

\textbf{\textit{AD-step:}} After acquiring the warped HR-MSI $\widehat{\textbf{Z}}$, we employ a asymmetric downsampling operation to reduce the spatial resolution of the image. The specific operation can be expressed as:
\begin{equation}
	\begin{aligned}
		\widetilde{\textbf{Z}} = &\, (\widehat{\textbf{Z}} * k)\downarrow_{s} \\
	\end{aligned}
	\label{eq:eq8}
\end{equation}
where $k$ represents the blur kernel, $*$ represents the convolution blur operation, and $\downarrow_{s}$ represents the interval sampling with factor $s$ $(s= \frac{H}{h}, s= \frac{W}{w})$. The spatially degraded HR-MSI $\widetilde{\textbf{Z}} \in \mathbb{R}^{h \times w \times c}$ is the output of SW-step and also the final output of SpaDN.

\begin{figure}[!t]
	\setlength{\abovecaptionskip}{1pt}
	\setlength{\belowcaptionskip}{-5pt}
	\begin{center}
		\includegraphics[scale=0.23]{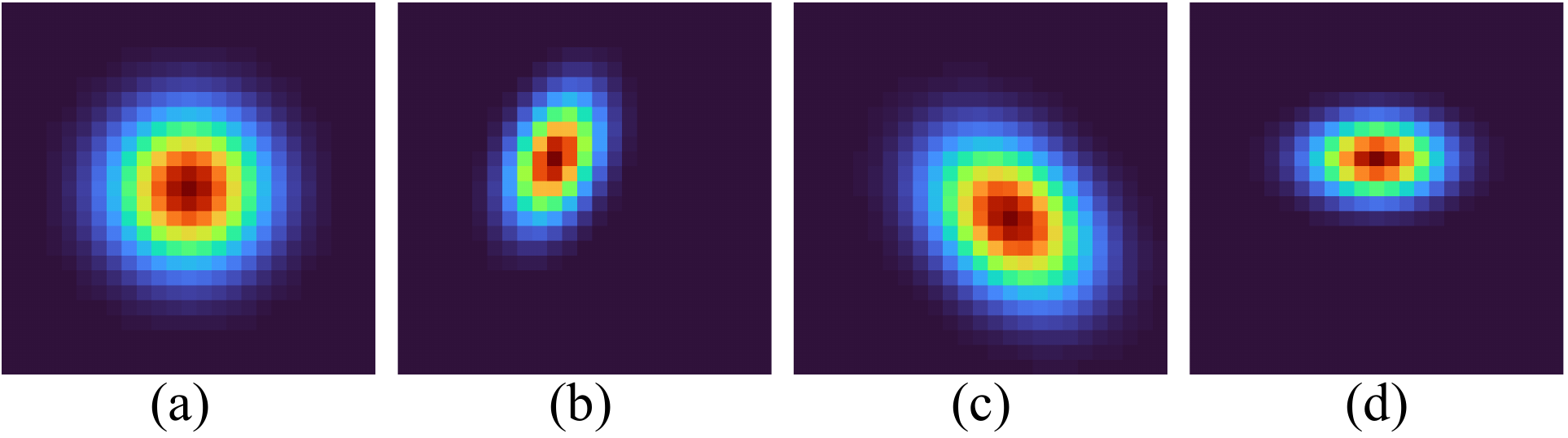}
	\end{center}
	\caption{Examples of generated shifted anisotropic Gaussian kernels. The parameters $[\alpha_y, \alpha_x, \delta_{y}, \delta_{x}, \beta]$ of these kernels are: (a) $[0, 0, 3, 3, 0]$. (b)$[-2, -2, 2.5, 1.5, \pi/8]$. (c) $[2, 2, 2.5, 3.5, \pi/4]$. (d) $[-2, 0, 1.5, 2.5, 0]$.}
	\label{fig:kernel}
\end{figure}

Notably, the term ``asymmetric" is manifested in that we propose a shifted anisotropic Gaussian blur kernel. This kernel can be mathematically expressed as:
\begin{equation}
	\begin{gathered}
		k_{i,j} = exp(-\frac{1}{2} \begin{bmatrix}i-\alpha_y \,\,\, j-\alpha_x\end{bmatrix}\mathrm{M}^{-1} \begin{bmatrix}i-\alpha_y\\j-\alpha_x\end{bmatrix}) \\
		s.t. \,\, \sum_{i,j} k_{i,j} = 1
	\end{gathered}
	\label{eq:eq9}
\end{equation}
where $i, j$ represent the coordinate of the kernel $k$, $\mathrm{M} \in \mathbb{R}^{2 \times 2}$ represents the covariance matrix, and $\alpha_y, \alpha_x$ represent the coordinate offsets along the y-axis and x-axis. $\mathrm{M}$ is used to control the shape of the kernel, which is formulated as: 
\begin{equation}
	\mathrm{M} =\begin{bmatrix}\delta_{y} & 0 \\ 0 & \delta_{x}\end{bmatrix} \begin{bmatrix} \cos\beta & -\sin\beta \\ \sin\beta & \cos\beta \end{bmatrix}
	\label{eq:eq10}
\end{equation}
where $\delta_{y}, \delta_{x}$ represent the standard deviations that control the stretch of the kernel $k$ along the y-axis and x-axis, and $\beta$ represents the angle that governs the rotation. As shown in Fig. \ref{fig:kernel}, the kernel of any shape can be parameterized as five learnable parameters, \ie, $\alpha_y, \alpha_x, \delta_{y}, \delta_{x}, \beta$.

In contrast to existing blur kernel modeling methods \cite{kernel1, kernel2}, our modeling approach accurately matches the physical characteristics of blurring. First, the data distribution characteristics of the Gaussian kernel highly match the physical essence of many blurring effects, where the central data intensity is maximum and gradually decreases towards the edges. Second, in practical scenarios, blurring often exhibits asymmetry due to various factors such as camera shake and atmospheric disturbances. Therefore, employing anisotropic stretching and coordinate shifting operations allows for a more accurate capture of these blurring patterns.

\subsection{Spectral Degradation Network}
\hspace{1em} SpeDN conducts spectral resolution reduction on the image, which corresponds to a physical transformation process that shifts the image from high to low spectral resolution, and the overall process displays non-uniformity. 
As shown in Fig. \ref{fig:pipeline}, the SpeDN divides this process into two implementation steps: the spectral modulation step (SM-step) and the parallel downsampling step (PD-step). Next, we introduce the details of SM-step and PD-step, respectively.

\textbf{\textit{SM-step:}}
The primary objective of SM-step is to introduce the non-uniformity into the spectral degradation process by simulating lens aberrations. Here, we apply a spectral modulation operation to achieve this objective:
\begin{equation}
	\widehat{\textbf{Y}} = f_{smn}(\texttt{concat}(\textbf{Y}, \Phi)) 
	\label{eq:eq11}
\end{equation}
where $\Phi \in \mathbb{R}^{h \times w \times D}$ represents the positional encoding matrix, $D$ represents the channel dimension of the positional encoding matrix $\Phi$, and $\texttt{concat}(,)$ represents the concatenation operation along the channel dimension. $f_{smn}$ represents the spectral modulator network that modulates the concatenated LR-HSI $\textbf{Y}$ and positional encoding matrix $\Phi$, and $\widehat{\textbf{Y}} \in \mathbb{R}^{h \times w \times C}$ is the modulated LR-HSI.

We encode position information into the positional encoding matrix $\Phi$, which is then concatenated with the LR-HSI $\textbf{Y}$ to establish a joint spatial-spectral representation. This operation enables the spectral modulator network $f_{smn}$ to simultaneously consider spatial and spectral information within a unified data representation, which allows for distinct modulations on pixels of the LR-HSI $\textbf{Y}$ at different locations. Additionally, certain locations might exhibit distinct spectral characteristics due to the factors like materials or illuminations. While the positional encoding matrix $\Phi$ can aid the spectral modulator network $f_{smn}$ in capturing this location-specificity.

Next, we present the procedure for procuring the positional encoding matrix $\Phi$. First, we record the coordinate information of the LR-HSI $\textbf{Y}$ into a grid matrix $\Lambda \in \mathbb{R}^{h \times w \times 2}$, where $h$ and $w$ denote the height and width of the LR-HSI $\textbf{Y}$, and the value 2 corresponds to the coordinate representation along the y-axis and x-axis. The specific coordinate representation $\Lambda_{i,j} = \begin{bmatrix} p_i \,\, q_j \end{bmatrix}$ is given by:
\begin{equation}
	\begin{aligned}
		p_i &= -1 + \frac{2(i - 1)}{h-1}, & i = 1, \dots, h \\
		q_j &= -1 + \frac{2(j - 1)}{w-1}, & j = 1, \dots, w
	\end{aligned}
	\label{eq:eq12}
\end{equation}
where the grid values are normalized to the range $\left[-1, 1\right]$ to facilitate the encoding operation.

Second, we propose a encoding matrix generator $f_{egen}$ to encode the grid matrix $\Lambda$, which is formulated as:
\begin{equation}
	\Phi = f_{egen}(\Lambda)
	\label{eq:eq13}
\end{equation}
Notably, the encoding operation can provide a unified high-dimensional feature space for grid matrix $\Lambda$ of different sizes, which enables the spectral modulator network $f_{smn}$ to have better scale invariance and be able to adapt to various sizes of inputs. Moreover, the grid matrix $\Lambda$ after dimensional enhancement can provide a richer description for each spatial position, which helps the spectral modulator network $f_{smn}$ understand the spatial distribution characteristics of spectral data during the modulation process.

\textbf{\textit{PD-step:}} Upon obtaining the modulated LR-HSI $\widehat{\textbf{Y}}$, we apply a parallel downsampling operation to reduce the spectral resolution of the image. This operation can be described as:
\begin{equation}
	\widetilde{\textbf{Y}} = \texttt{concat}(g_1(\widehat{\textbf{Y}}), g_2(\widehat{\textbf{Y}}), \dots, g_c(\widehat{\textbf{Y}}))
	\label{eq:eq14}
\end{equation}
where the spectrally degraded LR-HSI $\widetilde{\textbf{Y}} \in \mathbb{R}^{h \times w \times c}$ is the output of SM-step and also the final output of SpeDN. $\mathcal{G} = \{g_n\}^{c}_{n=1}$ represents a set of independent trainable band-focused blocks that execute parallel spectral degradations on the modulated LR-HSI $\widehat{\textbf{Y}}$, where $c$ represents the channel number of the spectral degraded LR-HSI $\widetilde{\textbf{Y}}$. Each block is specialized in generating a certain spectral band $\widetilde{y}_n \in \mathbb{R}^{h \times w \times 1}$ of the spectral degraded LR-HSI $\widetilde{\textbf{Y}}$ as:
\begin{equation}
	\begin{aligned}
		\widetilde{y}_n &= g_n(\widehat{\textbf{Y}}), & n = 1, \dots, c
	\end{aligned}
	\label{eq:eq15}
\end{equation}

We adopt the parallel strategy for the following reasons. First, spectral degradation refers to the conversion of high-dimensional spectral data into a lower-dimensional representation. In this process, it is essential to consider the physical characteristics of the spectral data, such as the absorption and reflection properties of objects for various wavelengths. Independent band-focused blocks allow for more detailed modeling of these physical characteristics for each spectral band. Second, in spectral data, noise levels can differ across wavelengths. Independent band-focused blocks have the capacity to adapt to the distinct noise distribution inherent in each spectral band.

\begin{table*}[!t]
	\setlength{\belowcaptionskip}{-5pt}
	\centering
	\hspace*{-0.3cm}
	\fontsize{9}{10}\selectfont
	\setlength{\tabcolsep}{12pt}
	\begin{tabular}{ccccc}
		\toprule
		Degradation Model & HypSen & WV & ZY & GF5-GF1 \\
		\midrule
		HyCoNet-DM & 0.4928/0.2107 & 0.8397/0.0320 & 0.7264/0.0731 & 0.8863/0.0302 \\
		MIAE-DM & 0.6233/0.1597 & 0.5199/0.1783 & 0.9781/0.0113 & 0.8327/0.1204 \\
		DFMF-DM & 0.6243/0.1500 & 0.8927/0.0476 & 0.9475/0.0240 & 0.9588/0.0212 \\
		ZSL-DM & 0.4821/0.1480 & 0.8393/0.0520 & 0.9053/0.0182 & 0.9502/0.0163 \\
		DAEM-DM & 0.4880/0.2143 & 0.4369/0.1934 & 0.6276/0.0840 & 0.6943/0.2090 \\
		BUSIF-DM & 0.5066/0.1878 & 0.7379/0.0758 & 0.7807/0.1170 & 0.8224/0.0466 \\
		\midrule
		PIDM (Ours) & \textbf{0.9431/0.0472} & \textbf{0.9766/0.0230} & \textbf{0.9928/0.0096} & \textbf{0.9834/0.0154} \\
		\bottomrule
	\end{tabular}
	\caption{Degradation modeling errors comparison (on SSIM and RMSE) on four real data. We add the word suffix ``-DM'' on the names of fusion methods to represent their respective degradation models.}
	\label{tab:ssim}
\end{table*}

Back to the band-focused block, there are two key components in each block, \ie, the GELU activation function and instance normalization (IN). Next, we  intend to explain why GELU and IN are chosen when building the band-focused block. \textbf{GELU:} Spectral data is typically continuous along the spectral bands, and the smooth non-linearity of GELU makes it well-suited to capture this continuity. In addition, spectral data often exhibits high dimensionality, and with its unsaturated, GELU can provide a more stable gradient flow, thereby preventing issues such as gradient vanishing or exploding in high-dimensional spaces. \textbf{IN}: In spectral data, different spectral bands may exhibit diverse intensity ranges and variations. IN ensures that each spectral band is appropriately normalized, which makes it easier for the block to capture and simulate degradation. 

The parallel strategy and band-focused block configurations aim to ensure the matching of the spectral resolution reduction process with the physical characteristics presented by the spectral data.

\section{Experiments}
\label{sec:experiments}
\hspace{1em} We evaluate PIDM under the TSG-mode and OFC-mode on four different real data to demonstrate the effectiveness and flexibility of our approach. Furthermore, to corroborate the authenticity of PIDM, additional tests are conducted using synthetic data.

\subsection{Datasets Description}
\hspace{1em}  \textit{1) Real Datasets Captured in Natural Scenarios:\footnote{We extend our gratitude to the authors of \cite{ual-hypsen, daem, cssnet-zy, dfmf} for providing the data. Their support is pivotal to our research.}:}
We collect four real datasets, including the HypSen \cite{ual-hypsen}, World View-2 (WV) \cite{daem}, Ziyuan 1-02D (ZY) \cite{cssnet-zy}, and Gaofen5-Gaofen1 (GF5-GF1) \cite{dfmf}. Each dataset comprises an LR-HSI and an HR-MSI, where the images within WV and GF5-GF1 are gathered from different satellites. After excluding noisy bands, selected subregions from the images are designated as the experimental data.

\textit{2) Auxiliary Datasets for Synthesizing Training Samples:}
We adopt the Pavia University \cite{hypconet}, Washington DC \cite{miae}, Chikusei \cite{fusion6} and Xiongan \cite{cssnet-zy} datasets as the benchmark for synthesizing training samples. More detailed information of these datasets is provided in the supplementary materials.

\subsection{Experiment Setup}
\hspace{1em} \textit{1) Comparison Methods:}
We select six state-of-the-art fusion methods for relevant experiments, each of which has made specific attempts in addressing the degradation modeling aspect. These methods include HyCoNet \cite{hypconet}, MIAE \cite{miae}, DFMF \cite{dfmf}, ZSL \cite{zsl}, DAEM \cite{daem}, and BUSIF \cite{busifusion}, which cover two usage modes of the degradation models. Specifically, ZSL and DFMF correspond to the TGS-mode, while HyCoNet, MIAE, DAEM, and BUSIF correspond to the OFC-mode. We first evaluate the degradation modeling errors of the degradation models across different fusion methods. Then, we assess the fusion performance of these fusion methods when equipped with PIDM.

\textit{2) Evaluation Metrics:} 
Following the works \cite{zsl, pan1}, we rely on the non-reference index QNR \cite{qnr} to quantitatively evaluate the quality of the fusion results. The QNR serves as a metric to assess the spectral and spatial quality of the fused HSI, with values ranging from 0 to 1. A higher QNR score indicates superior fusion results.

\textit{3) Implementation Details:} The training of PIDM is performed using PyTorch on an NVIDIA 3090 GPU employing an Adaptive Moment Estimation (ADAM) optimizer. The training process encompasses 2000 epochs, with a learning rate of 5e-4. We conduct training using the original image directly, without any patch-cutting operations. In the experiment, the width and height of the blur kernel $k$ are set to 25, the threshold $\epsilon$ of the truncation function $\tau_\epsilon(t)$ is set to 3, and the channel dimension $D$ of the positional encoding matrix $\Phi$ is set to 32. The source code will be released to be publicly available. For additional training details of various fusion methods and the clear network structures of our degradation models, please refer to the supplementary materials. 

\begin{figure*}[!t]
	\setlength{\abovecaptionskip}{-2pt}
	\setlength{\belowcaptionskip}{-1pt}
	\begin{center}
		\hspace*{-0.0cm}
		\includegraphics[scale=0.45]{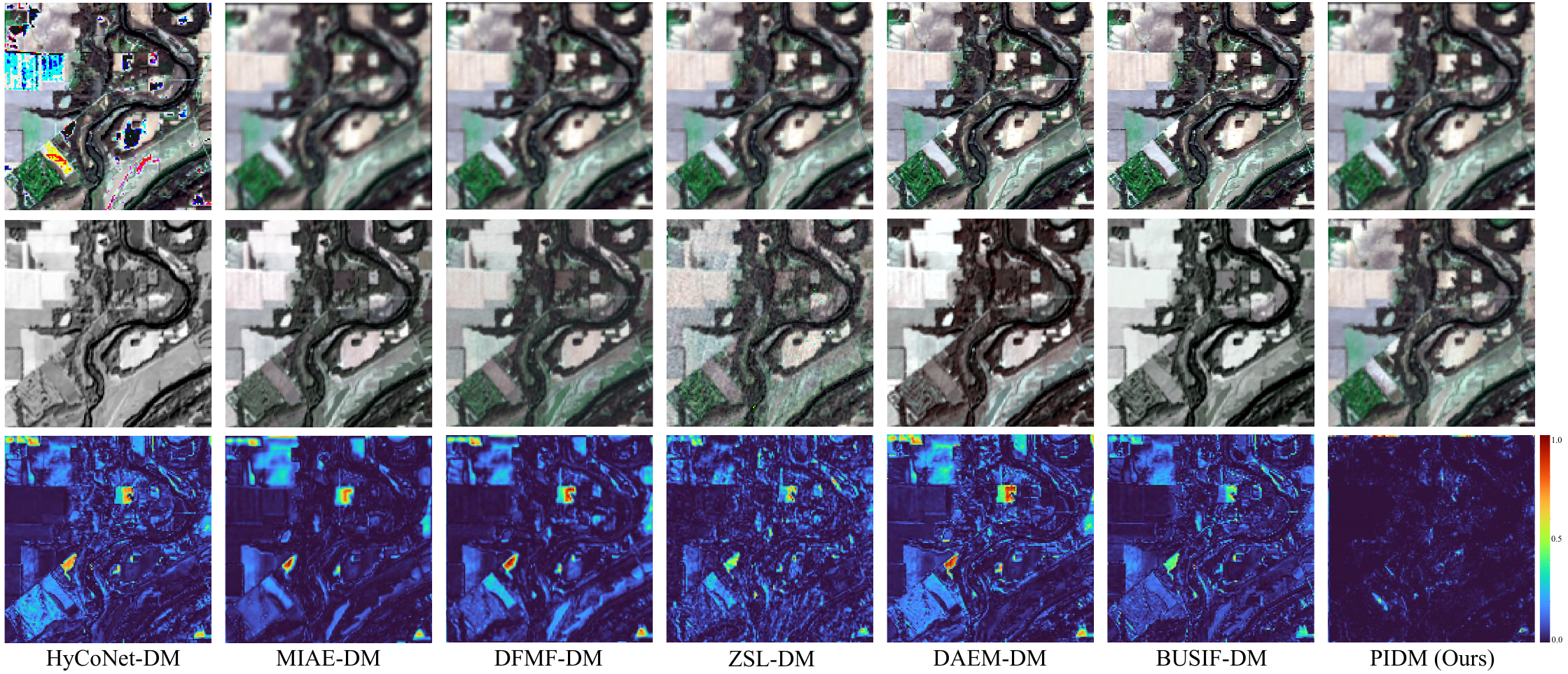}
	\end{center}
	\caption{Visualization (with bands 1-2-3 as B-G-R) of degradation estimation for each degradation model on the HypSen dataset. We add the word suffix ``-DM'' on the names of fusion methods to represent their respective degradation models. Row 1 represents the spatially degraded HR-MSI $\widetilde{\textbf{Z}}$ and row 2 represents the spectrally degraded LR-HSI $\widetilde{\textbf{Y}}$. Row 3 represents the modeling errors map.}
	\label{fig:esti}
\end{figure*}

\subsection{Comparison of Degradation Models}
\hspace{1em}  We believe that a more precise modeling on degradation should bring the spatially degraded HR-MSI $\widetilde{\textbf{Z}}$ and the spectrally degraded LR-HSI $\widetilde{\textbf{Y}}$ closer in terms of appearance. Thus, we compute the SSIM and RMSE metrics between the spatially degraded HR-MSI $\widetilde{\textbf{Z}}$ and the spectrally degraded LR-HSI $\widetilde{\textbf{Y}}$ to measure the degradation modeling errors in various fusion methods on four real data. The quantitative results are presented in Table \ref{tab:ssim}, and the visual results are shown in Fig. \ref{fig:esti}. Due to page limitation, we only present the visual results of the HypSen dataset, and more visual results can be found in the supplementary materials.

Table \ref{tab:ssim} shows that PIDM achieves the lowest degradation modeling errors across various real data. Especially on the HypSen dataset, PIDM far outperforms other degradation models. The HypSen dataset presents a highly intricate degradation scenario, while our PIDM can fully capture these complex degradation patterns. Visual results in Fig. \ref{fig:esti} further confirm the effectiveness of PIDM. In summary, PIDM can provide more accurate modeling for the degradation process in practical scenarios.

\subsection{Impact of Degradation Models on Fusion}
\label{sec:4.4}
\hspace{1em} In this section, we investigate whether more accurate degradation models can improve the performance of the fusion models on real data. We present the quantitative and visual results in Table \ref{tab:qnr} and Fig.\ref{fig:zy}, respectively. For lack of space, Fig.\ref{fig:zy} only illustrates the visual results of the ZY dataset, and more visual results are presented in the supplementary materials. The experimental results in Table \ref{tab:qnr} indicate that PDIM can enhance the performance of fusion methods, whether in the TGS-mode or OFC-mode. Visual results in Fig. \ref{fig:zy} also demonstrate that fusion methods can achieve better fusion details when employing our PIDM. In short, PIDM can boost the effectiveness of fusion methods in diverse practical scenarios and exhibits significant flexibility and adaptability.

\begin{table}[!t]
	\centering
	\setlength{\belowcaptionskip}{-5pt}
	\fontsize{9}{10}\selectfont
	\setlength{\tabcolsep}{4pt}
	\begin{tabular}{ccccc}
		\toprule
		Method & HypSen   &  WV   &  ZY   &  GF5-GF1 \\ \midrule
		HyCoNet &  0.8403   &  0.4421   &  0.9154   &  0.8506 \\
		HyCoNet w/ PIDM &  \textbf{0.9018}   &  \textbf{0.9295 }  &  \textbf{0.9178}   & \textbf{0.9581}   \\ \midrule
		MIAE &  0.8404   & 0.7208    &  0.8881   & 0.9601   \\
		MIAE w/ PIDM &  \textbf{0.8924 }  & \textbf{0.9561}    &  \textbf{0.9183}   & \textbf{0.9757}   \\ \midrule
		DFMF &  0.8641   & 0.9378    &  0.9576   &  0.9617  \\
		DFMF w/ PIDM &  \textbf{0.9325}   &  \textbf{0.9543}   &  \textbf{0.9601}   & \textbf{0.9620}   \\ \midrule
		ZSL &  0.7024   &  0.9507   &  0.7789   & 0.5039   \\
		ZSL w/ PIDM & \textbf{0.8709 }   &  \textbf{0.9699}   &   \textbf{0.8926}  & \textbf{0.8191}   \\ \midrule
		DAEM &  0.9034   &  0.9555   &  0.9159   & 0.9539  \\
		DAEM w/ PIDM & \textbf{0.9606}    &  \textbf{0.9649}   &  \textbf{0.9251}   &  \textbf{0.9635}  \\ \midrule
		BUSIF &  0.7659   &  0.8805   &  0.8982   &  0.7115  \\
		BUSIF w/ PIDM & \textbf{0.8504}    &   \textbf{0.9493}  &  \textbf{0.9205}   &  \textbf{0.9279} \\ \bottomrule
	\end{tabular}
	\caption{QNR scores across various fusion methods employing distinct degradation models on four real data. ``w/ PIDM'' means that the fusion method is equipped with our PIDM.}
	\label{tab:qnr}
\end{table}

\begin{figure*}[!t]
	\setlength{\abovecaptionskip}{0pt}
	\setlength{\belowcaptionskip}{0pt}
	\begin{center}
		\hspace*{-0.3cm}
		\includegraphics[scale=0.525]{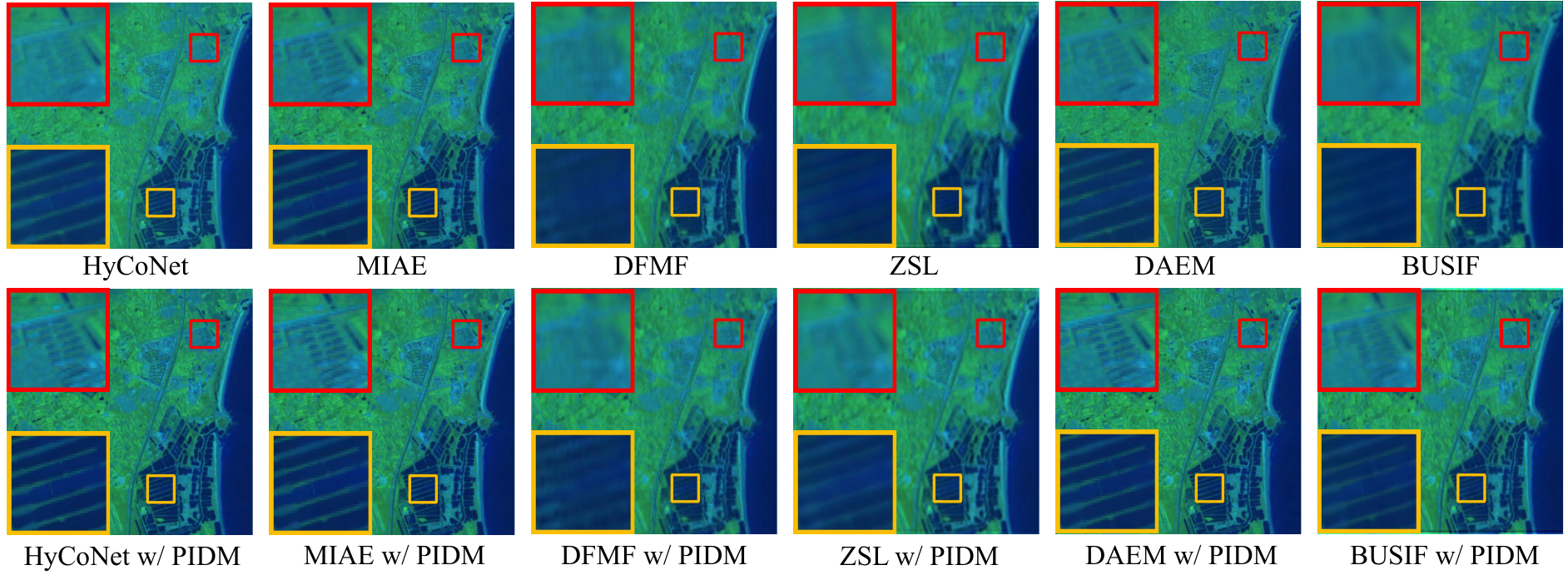}
	\end{center}
	\caption{Fusion results (with bands 20-40-60 as B-G-R) of various fusion methods employing distinct degradation models on the ZY dataset. ``w/ PIDM'' means that the fusion method is equipped with the PIDM.}
	\label{fig:zy}
\end{figure*}

\subsection{Effectiveness of Each Proposed Operation}
\hspace{1em} This section conducts an ablation study to evaluate the impact of parallel downsampling (PD), asymmetric downsampling (AD), spectral modulation (SM), and spatial warping (SW) operations. Firstly, we establish base degradation models that utilize a standard Gaussian kernel with size $25\times25$ and uniform downsampling for spatial degradation and adopt a trainable matrix for spectral degradation. Subsequently, we progressively modify the base degradation models by adding PD, AD, SM, and SW. Finally, various modified degradation models are incorporated into the fusion methods ZSL (TGS-mode) and BUSIF (OFC-mode) for fusion experiments. Ablation quantitative results on the WV dataset are reported in Table \ref{tab:abla}.

As shown in Table \ref{tab:abla}, the inclusion of PD, AD, SM, and SW leads to a continuous improvement in fusion results. This signifies the essential contribution of each operation in enhancing the accuracy of degradation modeling. Additionally, our modular approach exhibits high flexibility, which allows researchers to select and adapt specific operation to accommodate various degradation scenarios. 

\begin{table}[!t]
	\centering
	\setlength{\abovecaptionskip}{12pt}
	\setlength{\belowcaptionskip}{-6pt}
	\fontsize{9}{12}\selectfont
	\begin{tabular}{cccc|cc}
		\Xhline{0.7px}
		\multicolumn{4}{c}{Degradation Model Config} \vline & \multicolumn{2}{c}{Fusion Model} \\ \hline
		PD & AD & SM & SW & ZSL-FM    & BUSIF-FM  \\ \hline
		&              &              &            & 0.9507  & 0.7558     \\ 
		\checkmark      &              &          &           & 0.9513  & 0.8730     \\
		\checkmark      & \checkmark        &           &          & 0.9548  & 0.8811     \\
		\checkmark      & \checkmark      & \checkmark      &             & 0.9591  & 0.9400     \\
		\checkmark      & \checkmark      & \checkmark      & \checkmark     & \textbf{0.9681}  & \textbf{0.9493} \\   		\Xhline{0.7px}
	\end{tabular}
	\caption{QNR scores for the ablation study conducted with the fusion models in ZSL (ZSL-FM) and BUSIF (BUSIF-FM) on the WV dataset. Row 3 represents the experimental results of the base degradation models.}
	\label{tab:abla}
\end{table}

\subsection{Authenticity of Proposed Degradation Models}
\hspace{1em} We conduct synthetic data experiments to ascertain whether PIDM indeed captures the real-world degradation patterns. The Hypsen and ZY datasets and the fusion model in ZSL are chosen for the relevant experiments. Taking the row 3 in Table \ref{tab:synthetic} as an example, we first obtain the PIDM from HypSen and then apply PIDM to the auxiliary datasets to synthesize training samples. Subsequently, the ZSL fusion model is trained using the generated samples, and an evaluation of the ZSL fusion model is performed on HypSen. Notably, the number of spectral bands differs across the auxiliary datasets. Hence, we compress the band numbers of the auxiliary datasets to enable PIDM to generate training samples with band dimensions consistent with the real data. More details on experimental configurations are presented in the supplementary materials.

Table \ref{tab:synthetic} shows that the ZSL fusion model performs better when employing the PIDM for generating training samples. This indicates that PIDM indeed captures the real-world degradation patterns and can enhance the generalization ability of the fusion model by integrating these degradation patterns into the synthetic data.

\begin{table}[!t]
	\centering
	\setlength{\abovecaptionskip}{12pt}
	\setlength{\belowcaptionskip}{-6pt}
	\fontsize{9}{13}\selectfont
	\begin{tabular}{l|c|c}
		\Xhline{0.7px}
		Training Samples Generation &Testing Dataset & QNR \\ \hline
		Auxiliary + ZSL-DM &\multirow{2}{*}{HypSen} & 0.6184 \\
		Auxiliary  + PIDM & & 0.7560 \\ \hline
		Auxiliary  + ZSL-DM & \multirow{2}{*}{ZY} & 0.7333 \\
		Auxiliary  + PIDM & & 0.8597 \\ \Xhline{0.7px}
	\end{tabular}
	\caption{QNR scores of synthetic data experiments using the ZSL fusion model on the HypSen and ZY datasets. ``ZSL-DM'' represents the degradation models in the ZSL fusion method. }
	\label{tab:synthetic}
\end{table}

\section{Conclusion}
\label{sec:conclusion}
\hspace{1em}  In this paper, we propose PIDM, a novel approach that performs physics-inspired modeling for degradation processes. PIDM is designed based on two insights. First, PIDM incorporates non-uniformity into the degradation process by employing spatial warping and spectral modulation operations. Second, PIDM ensures the matching of degradation processes with specific physical characteristics by introducing asymmetric and parallel downsampling operations. Extensive experimental results demonstrate that PIDM exhibits significant flexibility and superiority, which provides a plug-and-play solution and can assist the fusion methods in achieving boosted fusion performance in practical scenarios. Our study introduces a new perspective on modeling degradation processes, and future work will develop a fusion algorithm that matches our physics-inspired degradation models.

{
	\small
	\bibliographystyle{ieeenat_fullname}
	\bibliography{main}
}


\end{document}